\documentclass[12pt,reqno]{amsart}
\newcommand{\Rset}{\mathbb{R}}

\newcommand{\rmd}{\operatorname{d}}

\newcommand{\tcP}{\tilde{\mathcal{P}}}
\newcommand{\cE}{\mathcal{E}}

\newcommand{\cC}{\mathcal{C}}
\newcommand{\cP}{\mathcal{P}}

\newcommand{\hW}{{\hat{W}}}
\newcommand{\hP}[1]{{\hat{P}^{(#1)}}}
\newcommand{\hPab}{\hP{\alpha,\beta}}
\newcommand{\hPa}{{\hat{P}^{(\alpha)}}}
\newcommand{\hL}[1]{{\hat{L}^{(#1)}}}
\newcommand{\hLk}{\hL{k}}

\newcommand{\hmuk}{{\hat{\mu}_{k}}}
\newcommand{\hmuab}{{\hat{\mu}_{\alpha,\beta}}}
\newcommand{\hWab}{{\hat{W}_{\alpha,\beta}}}

\theoremstyle{plain}
\newtheorem{thm}{Theorem}[section]

\newtheorem{prop}{Proposition}[section]
\newtheorem{lem}{Lemma}[section]

\theoremstyle{definition}
\newtheorem{definition}{Definition}[section]

\theoremstyle{remark}
\newtheorem{remark}{Remark}[section]

\begin{document}
\title[An extended class of orthogonal polynomials]{An extended
  class of orthogonal polynomials defined by a Sturm-Liouville problem}
\author{David G\'omez-Ullate}
\address{ Departamento de F\'isica Te\'orica II, Universidad Complutense de Madrid, 28040 Madrid, Spain}
\author{ Niky Kamran }
\address{Department of Mathematics and Statistics, McGill University
Montreal, QC, H3A 2K6, Canada}
\author{Robert Milson}
\address{Department of Mathematics and Statistics, Dalhousie University, Halifax, NS, B3H 3J5, Canada}
\begin{abstract}
  We present two infinite sequences of polynomial   eigenfunctions of a Sturm-Liouville problem. As opposed to the
  classical orthogonal polynomial systems, these sequences start
  with a polynomial of degree one.  We denote these polynomials as
  $X_1$-Jacobi and $X_1$-Laguerre and we prove that they are
  orthogonal with respect to a positive definite inner product defined
  over the the compact interval $[-1,1]$ or the half-line
  $[0,\infty)$, respectively, and they are a basis of the corresponding
  $L^2$ Hilbert spaces. Moreover, we prove a converse statement
  similar to Bochner's theorem for the classical orthogonal polynomial
  systems: if a self-adjoint second order operator has a complete set of polynomial eigenfunctions $\{p_i\}_{i=1}^\infty$, then it must be either
the $X_1$-Jacobi or the $X_1$-Laguerre Sturm-Liouville problem. A Rodrigues-type formula can be derived for both of the $X_1$ polynomial sequences.

\end{abstract}
\maketitle
\section{Introduction}

%

The classical orthogonal polynomial systems (OPS) of Hermite, Laguerre
and Jacobi  are most often characterized as the polynomial solutions
of a Sturm-Liouville problem, following the celebrated result by
S. Bochner: if an infinite sequence of polynomials
$\{P_n(x)\}_{n=0}^\infty$ satisfies a second order eigenvalue equation
of the form
\begin{equation}\label{eq:bochner}
p(x)P_n'' + q(x) P_n' + r(x) P_n(x)=\lambda_n P_n(x),\qquad n=0,1,2,\dots
\end{equation}
then $p(x),q(x)$ and $r(x)$ must be polynomials of degree $2,1$ and
$0$ respectively \cite{routh,Bo}.  In addition, if the
$\{P_n(x)\}_{n=0}^\infty$ sequence is an OPS, then it has to be (up to
an affine transformation of $x$) one of the classical orthogonal
polynomial systems of Jacobi, Laguerre or Hermite
\cite{Aczel,Mikolas,Feldmann,Lesky,KL97}

Much work has been done since the 1940s until present in different
generalizations and extensions of these classical families.  One main
line of research has dealt with polynomial sequences defined by
differential equations of order higher than two, leading to the
\emph{Bochner-Krall} orthogonal polynomial systems \cite{Krall}. For a
good review on this subject, see for instance \cite{EKLW}.

When the measure is supported over a discrete set, we speak of
discrete orthogonal polynomials. The equivalent to the classical
families (Meixner, Hahn, Kravchuk, Charlier, etc.) are orthogonal
polynomials that satisfy a difference equation of hyper-geometric
type instead of a differential equation. This topic is reviewed for
instance in \cite{IsvAs}.

Probably the most general class is that of the Askey-Wilson
polynomials, \cite{askey}, a generalization of the classical
families that satisfy  $q$-difference equations and reduce to the
classical and discrete families under special or limiting cases.
Corresponding generalizations of Bochner's theorem also exist for
polynomials in the Askey-Wilson scheme  \cite{grunbaum,ismail}.

Another possible generalization concerns the \emph{semi-classical}
orthogonal polynomials, characterized by the fact that the
log-derivative of the weight factor is a rational function
\cite{hendrikssen}. Semi-classical polynomials have similar properties
as their classical counterparts: they form a positive-definite
orthogonal family which is complete in the corresponding
$L^2(w)$-space, the sequence of their derivatives is not orthogonal
but quasi-orthogonal \cite{Ron79}, they satisfy a second order
differential equation of the form \eqref{eq:bochner}, where the
coefficients $p(x,n)$, $q(x,n)$ and $r(x,n)$ have an explicit
dependence on $n$ \cite{atkinson,Ron87}.  When the classical weights
are modified by multiplication by a rational function (with poles and
zeros outside the interval of orthogonality), the modified weights are
semi-classical \cite{RM89,atkinson}. Expressions for these orthogonal
polynomials can be obtained through the application of Uvarov's
determinantal formula \cite{U,ismail}.

In the present paper we introduce orthogonal polynomials with rational
weights that are eigenfunctions of a Sturm-Liouville operator and are
therefore fundamentally different from the semi-classical orthogonal
polynomials.  The application of Uvarov's determinantal formulas gives
rise to a sequence of polynomials that begins with the polynomials of
degree zero and consists of polynomials which are orthogonal to 1. In
contrast, the families described below start with a polynomial of
degree one and are not orthogonal to 1.  Our approach leads to novel
examples that are neither classical nor semi-classical.

Many of the generalizations referred to above aim to retain the nice
properties that derive from the Sturm-Liouville character of a
classical OPS. However, it seems to be a well established fact in the
literature that no complete orthogonal polynomial systems other than
the classical ones arise as solutions of a Sturm-Liouville problem.
This is indeed the case if the operator belongs to the Bochner class
\eqref{eq:bochner}, as was proved by Lesky \cite{Lesky}.

We argue that from the point of view of Sturm-Liouville theory this
restriction is not essential.  It has been observed \cite{elw04}
that certain instances of classical orthogonal polynomial families
have the following curious property: the polynomials are formal
eigenfunctions of the operator \eqref{eq:bochner}, but a finite
number of initial polynomials are not square integrable.  Consider,
for instance the family of Laguerre polynomials $P_n(x) =
L^{-1}_n(x)$, $n=0,1,2,\ldots$, or more specifically
\begin{align*}
  P_0(x) &= 1,\\
  P_1(x) &= \frac{1}{2} x(x-2),\\
  P_2(x) &=-\frac{1}{6} x(x^2-6x+6),\\
  \vdots\\
  P_n(x) &= -\frac{1}{n} x L^{(1)}_{n-1}(x),\quad n\geq 1;
\end{align*}
The orthogonality is with respect to the weight $W(x) dx = x^{-1}
e^{-x} dx$, which implies that $P_0(x)$ is not square integrable.
Only the polynomials $P_1, P_2, P_3,\ldots$ arise as eigenfunctions
of the corresponding Sturm-Liouville problem, and (therefore) it  is
this truncated sequence  which is complete in the $L^2(W(x)
dx,(0,\infty))$ space.

The following question is therefore of interest:
\begin{quote}
  \textit{What sequences of polynomials can arise as eigenfunctions of
    a Sturm-Liouville problem?}
\end{quote}
The main idea of our paper is to show that the answer to the above
question takes one outside the realm of classical and semi-classical
orthogonal polynomials.  In other words, if the sequence
$\{P_n\}_{n=m}^\infty$ is allowed to start with a degree $m\geq 1$
polynomial, then there exist \emph{complete} sequences of polynomial
eigenfunctions that obey differential equations different from
\eqref{eq:bochner}.


In this paper we treat the case $m=1$.  In Section \ref{sec:X1} we
introduce the $X_1$-Laguerre and $X_1$-Jacobi orthogonal polynomial
systems.  These novel families are crucial to our main result, Theorem
\ref{thm:main1} --- a classification of complete orthogonal polynomial
sequences \emph{starting with a degree one polynomial} that satisfy a
Sturm-Liouville problem.  This theorem can thus be viewed as the
corresponding extension of the classical results of Bochner and Lesky.

Completeness of the new polynomial families is proved in Section
\ref{sec:complete} using a
suitable extension of the Weierstrass
approximation theorem.  Section \ref{sec:proof} completes the proof
of the main theorem. Some of the results contained in this section
rest on the classification of \emph{exceptional polynomial
subspaces} of co-dimension one and the spaces of second order
differential operators which leave them invariant. We will use some
of these results without proof, referring the interested reader to
the recent publication \cite{GKMpart1} where all the details are
given.  Finally, Sections \ref{sec:propJ} and \ref{sec:propL}
describe some properties of the new polynomial families:
factorization  of the second order operator, Rodrigues-type formula,
normalization constants, relation with the classical families,
three-term recurrence relation and some basic properties of the
zeros.

By way of conclusion, we mention that since our paper was posted in
preprint form on the arXiv, the Schr\"odinger operators and
potentials for which our new orthogonal polynomials appear as
eigenfunctions (when multiplied by the corresponding weight) have
been determined and studied \cite{Q}. The potentials are
deformations of the radial oscillator or the Scarf I potential
obtained by the addition of rational functions, and they are shape
invariant.

\section{Definitions and Main results}
\label{sec:X1}

Orthogonal polynomial systems are usually understood to start with a
polynomial of degree $0$.  However, from the point of view of
Sturm-Liouville theory, this restriction is unnecessary.  The
preceding observation motivates the following.

\begin{definition}
  \label{def:pslp}
  We define a \emph{polynomial Sturm Liouville problem}, or PSLP for
  short, to be a self-adjoint Sturm-Liouville boundary value problem
  with a semi-bounded, pure-point spectrum and \emph{polynomial
    eigenfunctions}.
\end{definition}

\begin{definition}
  \label{def:kops}
  For integer $k\geq 0$, we will say that a polynomial sequence $\{
  y_n \}_{n=k}^\infty$ is degree $k$ ($k$-PS) if it starts with a
  polynomial of degree $k$ and $\deg y_n = n$.  A $k$-PS is a
  \emph{degree $k$ orthogonal polynomial system ($k$-OPS)} if there
  exists a positive measure $W(x)\, dx$ on an interval
  \begin{equation}
    \label{eq:Idef}
    I = (x_1,x_2),\quad -\infty\leq x_1<x_2\leq\infty
  \end{equation}
  such that
  \begin{itemize}
  \item[(i)] The moments are well defined:
    \begin{equation}
      \label{eq:finitemoments}
      \int_I x^n W(x)\, dx < \infty, \quad n=0,1,2,\ldots;
    \end{equation}
  \item[(ii)] The polynomials are orthogonal:
    \begin{equation}
      \label{eq:intorth}
      \int_I y_m(x) y_n(x) W(x) \, dx = 0,\quad m\neq n;
    \end{equation}
  \item[(iii)]The sequence is a basis for the Hilbert
    space  $L^2(I,W\, dx)$.
  \end{itemize}
\end{definition}
\noindent
We remark that:
\begin{enumerate}
\item[(i)] The assumption of self-adjointness in Definition
  \ref{def:pslp} means that the eigenfunctions of a PSLP form an OPS.
\item[(ii)] Item (iii) in  Definition \ref{def:kops} implies that a $0$-OPS
  cannot be a $k$-OPS for $k>0$.
\item[(iii)] The classical or
semi-classical systems are necessarily $0$-OPS. This means that the
$k$-OPS represent a genuinely new generalization of the classical
OPS, even if they share the property of admitting rational weights.
\end{enumerate}
\begin{definition}
  Consider a PSLP whose eigenfunctions form a $k$-OPS.  We call the
  polynomial system classical if the second-order differential
  equation in question is of Bochner type \eqref{eq:bochner}.
  Otherwise we call the polynomial system \emph{exceptional}, or $X_k$
  for short.
\end{definition}
\noindent It is known that classical Laguerre polynomials with
negative integer parameters constitute a $k$-OPS where $k\geq 1$
\cite{elw04}. Some partial results are also available for Jacobi
polynomials with negative integer parameters \cite{amr02}. We
believe these to be the only classical examples where the polynomial
eigenfunctions of a second-order Sturm-Liouville problem begin with
a degree $k\geq 1$, but to our best knowledge this question has not
been explicitly investigated in the literature.  Turning to
exceptional polynomial families, Bochner's result shows that an
$X_0$ polynomial system is impossible.  By contrast, the $X_1$
definition is non-vacuous.

\subsection{$X_1$-Jacobi polynomials}
Let $\alpha\neq \beta$ be real parameters and
\begin{subequations}
  \label{eq:abcfromalphabeta}
\begin{equation}\label{eq:abdef}
  a= \frac{1}{2}(\beta-\alpha),\quad
  b= \frac{\beta+\alpha}{\beta-\alpha},\quad
\end{equation}
\begin{equation}
  c=b+1/a.\label{eq:cdef}
  \end{equation}
\end{subequations}
Consider the polynomials
\begin{equation}\label{eq:ui} u_1=x-c,\qquad u_i=(x-b)^i,\quad
  i\geq 2,\end{equation}
the first $n$ of which provide a basis of the space $\cE_n^{a,b}$:
\begin{eqnarray}\cE_n^{a,b}&\equiv&\{p\in\cP_n\,|\, p'(b)+ap(b)=0\} \label{eq:Eabdef}\\
&=&\operatorname{span}\{u_1,u_2,\dots,u_n\}
\end{eqnarray}
The following restrictions will be required on the real parameters $\alpha,\beta$:
\begin{subequations}\label{alfabetacond}
\begin{eqnarray}
&&\alpha>-1,\quad \beta>-1, \label{alfabetacond1}\\
&&\text{sgn} \alpha=\text{sgn}\beta, \label{alfabetacond2}
\end{eqnarray}
\end{subequations}
the last of which ensures $|b|>1$.
We define the following measure
\begin{eqnarray}
   d\hmuab&=&\hWab\,dx, \quad x\in (-1,1),\\
 \hWab &=& \frac{(1-x)^\alpha  (1+x)^\beta}{(x-b)^2}=\frac{(1-x)^\alpha  (1+x)^\beta}{(x-\frac{\beta+\alpha}{\beta-\alpha})^2},
   \label{eq:Wabdef}
\end{eqnarray}
and observe that $\hWab>0$ for $-1<x<1$ so the scalar product
\begin{equation}
  \label{eq:jacobiproduct}
   (f,g)_{\alpha,\beta} := \int^1_{-1} f(x) g(x)\,  d\hmuab,
\end{equation}
is positive definite.
\begin{definition}
We define the $X_1$-Jacobi polynomial sequence
$\left\{\hPab_i\right\}_{i=1}^\infty$ as the polynomials obtained by
Gram-Schmidt orthogonalization from the sequence
$\{u_i\}_{i=1}^\infty$ in \eqref{eq:ui} with respect to the scalar
product \eqref{eq:jacobiproduct}, and by imposing  the normalization
condition
\begin{equation}
  \label{eq:jacobinorm}
  \hPab_n(1) = \frac{\alpha+n}{(\beta-\alpha)}\,
  \binom{\alpha+n-2}{n-1}.
\end{equation}
\end{definition}
\noindent
From their definition it is obvious that $\deg \hPab_n=n$. However, as
opposed to the ordinary Jacobi polynomials, the sequence starts with a
degree one polynomial.

\subsection{$X_1$-Laguerre polynomials}

Let $k>0$ be a real parameter. Similarly, consider now the sequence
\begin{equation}
  \label{eq:vi}
  v_1=x+k+1,\qquad v_i=(x+k)^i,\quad
  i\geq 2
\end{equation}
We define the following measure on the interval $x\in(0,\infty)$:
\begin{eqnarray}
d\hmuk&=& \hat W_k\,dx,\\
  \label{eq:Wkdef}
 \hat W_k &=& \frac{e^{-x} x^k}{(x+k)^2},
 \end{eqnarray}
and observe that $\hat W_k>0$ on the domain in question so the following scalar product is positive definite:
\begin{equation}
  \label{eq:Laguerreproduct}
   (f,g)_k:= \int^\infty_{0} f(x) g(x)\,d\hmuk,
\end{equation}
\begin{definition}
  We define the $X_1$-Laguerre polynomial sequence
  $\left\{\hLk_i\right\}_{i=1}^\infty$ as the polynomials
  obtained by Gram-Schmidt orthogonalization from the sequence
  $\{v_i(x)\}_{i=1}^\infty$ in \eqref{eq:vi} with respect to the
  scalar product \eqref{eq:Laguerreproduct} and subject to the
  normalization condition
  \begin{equation}
    \label{eq:lagnorm}
    \hLk_n(x) = \frac{(-1)^nx^n}{(n-1)!} + \text{ lower order terms}\, n\geq 1.
  \end{equation}
  Note that the $X_{1}$-Laguerre polynomial sequence starts with a
  polynomial of degree 1.
\end{definition}

\begin{definition}
  For $\alpha,\beta$ subject to the restrictions \eqref{alfabetacond}, let
  \begin{equation}
    \label{eq:Tabdef}
    T_{\alpha,\beta}(y) = (x^2-1) y''+2a
    \left(\frac{1-b\,x}{b-x}\right) \big((x-c)y'-y\big),
  \end{equation}
  where $a$, $b$ and $c$ are related to $\alpha,\beta$ by
  \eqref{eq:abcfromalphabeta}.  We define the $X_1$-Jacobi
  boundary value problem to be the differential equation
  \begin{subequations}
    \label{eq:JacobiSLP}
    \begin{equation}
      \label{eq:JacobiSLPeq}
      T_{\alpha,\beta}(y)  = \lambda y,
    \end{equation}
    where $y=y(x)$ is a twice-differentiable function defined on $x\in
    (-1,1)$ subject to the boundary conditions
    \begin{align}
      \label{eq:JacobiSLPa}
      &\lim_{x\to 1^-} (1-x)^{\alpha+1} (y(x)-(x-c)y'(x)) = 0,\\
      \label{eq:JacobiSLPb}
      &\lim_{x\to -1^+} (1+x)^{\beta+1} (y(x)-(x-c)y'(x)) = 0.
    \end{align}
  \end{subequations}
\end{definition}
\begin{definition}
  For $k> 0$ let
  \begin{equation}
    \label{eq:Tkdef}
    T_k(y)= -x   y''+\left(\frac{x-k}{x+k}\right)\big((x+k+1)y'-y\big)
  \end{equation}
  We define the $X_1$-Laguerre boundary value problem to be the
  differential equation
  \begin{subequations}
    \label{eq:LaguerreSLP}
    \begin{equation}
    T_k(y) = \lambda y,
    \end{equation}
    where $y=y(x)$ is a twice differentiable function on $x\in
    (0,+\infty)$ subject to the boundary conditions
    \begin{align}
      &\lim_{x\to 0^+} x^{k+1} e^{-x} (y(x)-(x-c) y'(x))=0,\\
      & \lim_{x\to \infty} x^{k+1}e^{-x} (y(x)-(x-c)y'(x)) = 0.
    \end{align}
  \end{subequations}
\end{definition}




\noindent
We are now ready to state the main result of this paper.
\begin{thm}
  \label{thm:main1}
  The $X_1$-Jacobi and $X_1$-Laguerre boundary value problems are
  PSLPs. Their respective eigenfunctions are the $X_1$-Jacobi and
  $X_1$-Laguerre 1-OPSs defined above;  we have,
  \begin{align}\label{eq:Jacobidiffeq}
    T_{\alpha,\beta}\hPab_n&=(n-1)(\alpha+\beta+n)\, \hPab_n, &&
    n=1,2,\dots,\\
    \label{eq:Laguerrediffeq}
    T_k\hLk_n&=(n-1)\, \hLk_n, && n=1,2,\dots
  \end{align}
  Conversely, if all the eigenpolynomials of a PSLP form a 1-OPS, then
  up to an affine transformation of the independent variable, the
  family
  in question is either a classical 1-OPS, or $X_1$-Jacobi, or
  $X_1$-Laguerre.
\end{thm}

\noindent
At this point some remarks are due in turn:
\begin{enumerate}
\item[i)] Observe that although the components of
  $T_{\alpha,\beta}$ and $T_k$ are \emph{rational} functions, these
  operators have an infinite family of \emph{polynomial} eigenfunctions.
\item[ii)] Note that both equations belong to the Heine-Stieltjes class \cite{He,St}, i.e. they can be written as $p y''+q y ' + r y=0$ where $p,q$ and $r$ are polynomials of degrees $3,2$ and $1$ respectively.
\item[iii)] The existence of these new families of polynomial
  eigenfunctions of a second order eigenvalue equation is not in
  contradiction with Bochner's theorem, since one of its premises is
  that the countable sequence of polynomial eigenfunctions should
  begin with a constant.
\item[iv)] Since the sequences start with a first degree polynomial,
  one might think at first that they cannot be dense in the
  corresponding $L^2$ space, but we shall see below that this is not
  the case.
\item[v)] The differential expression \eqref{eq:Tabdef} defines a
  unbounded operator on a suitably chosen dense subspace of
  $L^2((-1,1),d\hmuab)$.  As per the general Sturm-Liouville theory
  \cite{EverittJL}, if one takes the maximal such domain and restricts
  it by imposing boundary conditions \eqref{eq:JacobiSLPa}
  \eqref{eq:JacobiSLPb}, one obtain a self-adjoint operator.
  Alternatively, one can construct a self-adjoint operator by showing
  that an operator with $\cE^{a,b}$ as the domain is essentially
  self-adjoint.  This approach is carried out in Section
  \ref{sec:proof}.  Similar remarks hold for the Laguerre case.
\item[vi)]  Both the X1-Jacobi and the X1-Laguerre SLPs admit
  limit-point and limit-circle subcases depending on the value of the
  parameters $\alpha,\beta, k$.  Details of this analysis can be found
  in \cite{EverittJL}.
\end{enumerate}

The proof of Theorem \ref{thm:main1} is based on the classification of
$X_1$ subspaces. For this reason, we recall the necessary results and
definitions of this classification below, referring the reader to
\cite{GKMpart1} for further details and proofs.
\begin{definition}
  Let $M$ be an $n$-dimensional subspace of
  \[\cP_n=\operatorname{span}\{1,x,x^2,\dots,x^n \}.\]
  We say that $M$ is a \emph{codimension one exceptional subspace} of $\cP_n$
  (\emph{$X_1$-subspace}), if there exists a second order differential
  operator $T$ such that $TM\subset M$ but $T\cP_n\not\subset\cP_n$.
\end{definition}
The main result of the classification of $X_1$ spaces performed in \cite{GKMpart1} states that every $X_1$-space is projectively equivalent to the space $\cE_n^{a,b}$ defined in \eqref{eq:Eabdef}. For the scope of this study we shall require a stronger property, namely that the differential operator $T$ preserves each subspace of the infinite flag
\begin{eqnarray}
&&\cE_1^{a,b}\subset \cE_2^{a,b}\subset \cE_3^{a,b}\subset \cdots\\
&&T \cE_n^{a,b}\subset \cE_n^{a,b},\qquad \forall n\geq1.
\end{eqnarray}

Let $a,b,c$ be real constants related by \eqref{eq:cdef} and set
\begin{subequations}\label{eq:Tpqrdef}
\begin{align}
  \label{eq:pxdef}
  p(x)&=k_2 (x-b)^2+ k_1 (x-b) + k_0, \\
  \label{eq:tqdef}
  \tilde{q}(x) &= a(x-c)(k_1 (x-b)+ 2k_0),\\
  \label{eq:trdef}
  \tilde{r}(x)&= -a(k_1 (x-b)+2 k_0),
\end{align}
\end{subequations}
where $k_0,k_1$ and $k_2$ are real constants and we assume that $k_0\neq 0$.

Let $T$ define the second-order operator
\begin{equation}
  \label{eq:TX1def}
  T(y):= p(x)y'' + \frac{\tilde{q}(x)}{x-b} y'
  +\frac{\tilde{r}(x)}{x-b} \, y.
\end{equation}
We are now ready to state the following theorem whose proof can be found in \cite{GKMpart1}:
\begin{thm}
  \label{thm:part1main}
  The operator $T$ defined in \eqref{eq:TX1def} with \eqref{eq:Tpqrdef} leaves invariant
  $\cE^{a,b}_n$ for all $n\geq 1$. Therefore, the eigenvalue equation
  \begin{equation}
    \label{eq:eigenproblem}
    Ty_n = \lambda_n y_n
  \end{equation}
  defines a sequence of polynomials $\{ y_n(x)\}_{n=1}^\infty$, where
  $y_n \in \cE^{a,b}_n$ with
  $n=\deg y_n$ and where
  \begin{equation}
    \label{eq:lambdandef}
    \lambda_n =  (n-1)(nk_2+a k_1),\quad n\geq 1.
  \end{equation}
  Conversely, suppose that $T$
  is a second-order differential operator such that the eigenvalue
  equation \eqref{eq:eigenproblem} is satisfied by polynomials
  $y_n(x)$ for all degrees $n\geq1$, but not for $n=0$.  Then, up
  to an additive constant, $T$ has the form \eqref{eq:TX1def} subject
  to \eqref{eq:Tpqrdef}, and $y_n \in \cE^{a,b}_n$.
\end{thm}

\begin{remark}\label{rem:X1}
The $X_1$-Jacobi and $X_1$-Laguerre
operators defined in \eqref{eq:Tabdef} and \eqref{eq:Tkdef} are
particular instances of the general $X_1$ operator \eqref{eq:TX1def} with \eqref{eq:Tpqrdef}.
In particular, for the $X_1$-Jacobi take $p(x)=x^2-1$ and the parameters
$\alpha,\beta$ are related to $a,b,c$ by \eqref{eq:abcfromalphabeta}.
For the $X_1$-Laguerre take $p(x)=-x$ and
\begin{equation}
  \label{eq:kabdef}
  a=-1,\quad b=-k,\quad c=-(k+1).
\end{equation}
With the choices above, the general eigenvalue formula \eqref{eq:lambdandef} provides the spectrum of the $X_1$-Jacobi and $X_1$-Laguerre operators in \eqref{eq:Jacobidiffeq} and \eqref{eq:Laguerrediffeq}.
\end{remark}

\section{Completeness of the $X_1$-Jacobi and $X_1$-Laguerre polynomial sequences}\label{sec:complete}

In this section we establish the completeness of the $X_1$-Jacobi and $X_1$-Laguerre polynomial sequences in their corresponding $L^2$ spaces. This fact might at first seem counter-intuitive since the classical polynomial sequences are no longer complete if the constants are removed from the sequence.

Before we prove this result, it is convenient to state the following
useful lemma, essentially a trivial extension of Weierstrass
approximation theorem, which can also be applied to higher codimension
polynomial subspaces.
\begin{lem}\label{lemma1}
  Let $\cP$ denote the ring of polynomials in $x\in\mathbb R$ with
  real coefficients and define $\tcP\subset\cP$ to be the following
  subspace of $\cP$:
\[
 \tcP=\left\{ p\in\cP \mid \sum_{j=0}^{r_i}a_{ij}p^{(j)}(x_i)=0,\quad i=1,\dots,k. \right\}
\]
where the $k$ points $x_1,\dots,x_k\,\notin[-1,1]$ and $p^{(j)}(x_i)$
denotes the $j$-th derivative of $p$ evaluated at $x_i$.

Then $\tcP$ is dense in $\cC[-1,1]$ with respect to the supremum norm.
\end{lem}
Note that the previous lemma also holds if two or more points $x_i$ are allowed to coincide, i.e. if more than one condition is imposed at each point.
\begin{proof}

We need to show that given an arbitrary $f\in\cC[-1,1]$ and any $\epsilon>0$, there exists a polynomial $\tilde p\in\tcP$ such that
\[ | f(x)-\tilde p(x)|<\epsilon\quad \forall x\in[-1,1]. \]
Consider the function
\[
 g(x)=\frac{f(x)}{\prod_{i=1}^{k}(x-x_i)^{1+r_i}}\in\cC[-1,1] \]
since all the poles $x_i$ lie outside the interval $[-1,1]$. By the Weierstrass approximation theorem, there exists a polynomial $p\in\cP$ such that
\[|g(x)-p(x)|<\frac{\epsilon}{\alpha}\quad \forall x\in[-1,1],\qquad \text{where } \alpha=\prod_{i=1}^k (1+|x_i|)^{1+r_i}\]
But then, the polynomial $\tilde p=\prod_{i=1}^{k}(x-x_i)^{1+r_i}\,p(x)$ belongs to $\tcP$ and we have
\[|f(x)-\tilde p(x)|=\left|\prod_{i=1}^{k}(x-x_i)^{1+r_i}\right|\cdot|g(x)-p(x)|<\epsilon \quad \forall x\in[-1,1]\]
since $|(x-x_i)^{1+r_i}|<(1+|x_i|)^{1+r_i}$ for $x\in[-1,1]$.
\end{proof}

\begin{prop}\label{prop:Eab}
If $|b|>1$, the space $\cE^{a,b}=\bigcup _n \cE^{a,b}_n$ is dense in $L^2([-1,1],\hWab)$.
\end{prop}
\begin{proof}
Since
\[ \cE^{a,b} = \left\{ p\in \cP\,|\, p'(b)+ap(b)=0 \right\}, \]
and $|b|>1$, Lemma \ref{lemma1} ensures that $ \cE^{a,b}$ is dense in $C[-1,1]$ with respect to the supremum norm, therefore also dense in $L^2([-1,1],\hWab)$.
\end{proof}
\begin{prop}\label{prop:X1JacOPS}
The $X_1$-Jacobi polynomial sequence $\left\{\hPab_i\right\}_{i=1}^\infty$ is a $1$-OPS.
\end{prop}

\begin{proof}
The sequence $\left\{\hPab_i\right\}_{i=1}^\infty$ is orthogonal by construction, it suffices then to prove that it is a basis of  $L^2([-1,1],\hWab)$. But by definition $\operatorname{span}\left\{\hPab_i\right\}_{i=1}^\infty=\cE^{a,b}$, so Proposition \ref{prop:Eab} states the desired result.
\end{proof}

In order to prove that the $X_1$-Laguerre polynomials $\{\hLk_i\}_{i=1}^\infty$  are an orthogonal basis of $L^2([0,\infty),\hmuk)$  we cannot use Lemma \ref{lemma1} since it only applies to the compact case. To this end, we state and prove the following:
\begin{lem}\label{lemma2}
The vector space
\[\tilde \cE=\left\{ p\in \cP\,|\, p(-k)=0 \right\}\]
is dense on the Hilbert space $L^2([0,\infty),\mu_k)$ where
\[d\mu_k=(x+k)^2 d\hmuk=x^k {\rm e}^{-x}\,dx.\]
\end{lem}

\begin{proof}
  Since $\cP=\Rset\oplus\tilde \cE$ it suffices to show that $1$ is in the $L^2(\mu_k)$-closure of $\tilde \cE$.
  To that end define the function
  \[ f(x)   \begin{cases}
    0 & \text{ if } 0\leq x<k,\\
    1/x & \text{ if } x\geq k.
  \end{cases}
  \]
  which is clearly in $L^2([0,\infty),\mu_k)$. Since the associated Laguerre polynomials are dense in $L^2([0,\infty),\mu_k)$, \cite{Sz}, there exists a polynomial $p\in\cP$ such that
  \[ \int^\infty_0 |f(x) - p(x)|^2 x^{k+2} {\rm e}^{-x}\, dx < {\rm e}^{-k} \epsilon,\]
for a given $\epsilon>0$.  Hence,
  \begin{eqnarray*}
    \int^\infty_0& |1-(x+k)&p(x+k)|^2 x^k {\rm e}^{-x} \,dx =\\
&=& \int^\infty_0 | 1/(x+k) - p(x+k)|^2 (x+k)^2 x^k {\rm e}^{-x} \, dx\\
    & \leq&   \int^\infty_0 | 1/(x+k) - p(x+k)|^2 (x+k)^{k+2} {\rm e}^{-x} \, dx\\
    & =&   {\rm e}^k\int^\infty_k | 1/x - p(x)|^2 x^{k+2} {\rm e}^{-x} \, dx\\
    & \leq& {\rm e}^k \int^\infty_0 | f(x) - p(x)|^2 x^{k+2} {\rm e}^{-x} \, dx\\
    & \leq& \epsilon
  \end{eqnarray*}
\end{proof}
We can now prove the following
\begin{prop}\label{prop:Laguerredense}
The $X_1$-Laguerre polynomials $\left\{\hLk_i\right\}_{i=1}^\infty$ are an orthogonal basis of $L^2([0,\infty),\hmuk)$.
\end{prop}

\begin{proof}
Since $\left\{\hLk_i\right\}_{i=1}^\infty$ are defined by Gram-Schmidt orthogonalization from the sequence $\{v_i(x)\}_{i=1}^\infty$, the set is orthogonal by construction and it suffices then to prove that
\[
\cE^{-1,-k}:=\text{span} \{v_i\}_{i=1}^\infty \text{ is dense in } L^2([0,\infty),\hmuk).
\]
Given an arbitrary $f\in L^2([0,\infty),\hmuk)$ and $\epsilon>0$, set
\[ \tilde{f}(x) = f(x)/(x+k),\quad x\geq 0,\]
and note that $\tilde{f} \in L^2([0,\infty),\mu_k)$.
Lemma \ref{lemma2} ensures that a polynomial $p(x)$ exists such that
\[ \int^\infty_0 | \tilde{f}(x) - (x+k) p(x)|^2 x^k {\rm e}^{-x}\, dx <
\epsilon,\]
Therefore
\[ \int^\infty_0 |f(x) - (x+k)^2 p(x)|^2 d\hmuk <\epsilon\]
and since $(x+k)^2 p(x)\in \cE^{-1,-k}$ this completes the proof.
\end{proof}

\section{Proof of Theorem \ref{thm:main1}}\label{sec:proof}

We begin by recalling some basic facts of Sturm-Liouville theory. An
arbitrary second order eigenvalue equation
\begin{eqnarray*}
 T(y)&=&p(x) y'' + q(x) y' + r(x) y,\\
 T(y)&=&\lambda y, \end{eqnarray*}
can be written in self-adjoint form
\[ ((p W y')'(x) + r(x) W(x) y(x)= \lambda W(x) y(x),\]
provided the function $W(x)$ satisfies a Pearson's type first order equation
\begin{equation}
  \label{eq:pearson}
  (p(x) W(x))' - q(x) W(x) = 0.
\end{equation}
which determines $W(x)$ uniquely up to a multiplicative factor as
\begin{equation}
  \label{eq:Wintform}
  W(x) =p(x)^{-1} \exp\left(\int^x \frac{ q(\xi)}{p(\xi)} \rmd\xi\right).
\end{equation}
The following well-known identity establishes the formal
self-adjointness of $T$ relative to the measure $W(x) dx$:
  \begin{eqnarray}
   \label{eq:Green} \int^{x_2}_{x_1} (T(f)g - T(g)f)(x) W(x)\, dx  =&&\\
    && \!\!\!\!\!\!\!\!\!\!\!\!\!\!\!\!\!\!\!\!\!\!\!\!\!\!\!\!\!\!\!\!\!\!\!\!\!\!\!\!\!\!\!\!\!=\Big[p(x) W(x) ( f'(x) g(x) - f(x) g'(x))\Big]^{x_2}_{x_1}.
\nonumber
\end{eqnarray}
where $-\infty\leq x_1<x_2\leq\infty$ and $f(x), g(x)$ sufficiently
  differentiable functions. The operator $T$ is symmetric if boundary conditions are imposed such that the right hand side of \eqref{eq:Green} vanishes.
If $y_1, y_2$ satisfy the eigenvalue equation
\[ T y_i = \lambda_i y_i,\quad i=1,2\]
with $\lambda_1 \neq \lambda_2$ and $T$ is symmetric, we
have
\[ (\lambda_1-\lambda_2) \int^{x_2}_{x_1} y_1(x) y_2(x) W(x) \, dx  = 0 .\]
so $y_1, y_2$ are orthogonal relative to $W(x)\, dx$.
\begin{remark}\label{rem:weights}
The weight function $\hWab$ defined in \eqref{eq:Wabdef} satisfies Pearson's equation
\eqref{eq:pearson} for
$T=T_{\alpha,\beta}$
shown in \eqref{eq:Tabdef}. Similarly, the weight function $\hW_k$ defined in
\eqref{eq:Wkdef} satisfies \eqref{eq:pearson} for $T=T_k$ defined in
\eqref{eq:Tkdef}.
\end{remark}

\subsection*{Forward statement of Theorem \ref{thm:main1}}
We can now prove the forward implication of Theorem \ref{thm:main1},
namely that the $X_1$-Jacobi and $X_1$-Laguerre SLPs defined in
\eqref{eq:JacobiSLP} and \eqref{eq:LaguerreSLP} have a simple, pure-point
spectrum bounded from below and a 1-PS of eigenfunctions.

Let us argue the $X_1$-Jacobi case and observe that the same arguments
apply \emph{mutatis mutandis} to the $X_1$-Laguerre case.  Consider
the operator $T_{\alpha,\beta}$ in \eqref{eq:Tabdef} defined on the
domain $\cE^{a,b}$. We will show that $T_{\alpha,\beta}$ is
essentially self-adjoint. By Theorem \ref{thm:part1main} and Remark
\ref{rem:X1} there exist polynomial eigenfunctions $y_n \in
\cE^{a,b}_n,$ $n\geq1$, for which \eqref{eq:JacobiSLPeq} holds. Since
$y_n$ satisfy the boundary conditions of the SLP \eqref{eq:JacobiSLP},
Remark \ref{rem:weights} and Green's identity \eqref{eq:Green} imply
that $\{ y_n\}_{n=1}^\infty$ are orthogonal with respect to the weight
$\hWab$ in \eqref{eq:Wabdef}. Moreover, $T_{\alpha,\beta}$ is a
symmetric and semi-bounded operator, so it must have a self-adjoint
extension $\tilde T_{\alpha,\beta}$ (see section X.3 in
\cite{RS}). All the eigenfunctions $y_n$ of $T_{\alpha,\beta}$ are
also eigenfunctions of $\tilde T_{\alpha,\beta}$ and Proposition
\ref{prop:Eab} states that $\{ y_n\}_{n=1}^\infty$ is a basis of
$L^2([-1,1],\hWab)$. Therefore, the resolution of the id entity
associated to $\tilde T_{\alpha,\beta}$ contains an infinite sum over
the corresponding projectors, and we conclude that the spectrum is
discrete and bounded from below, and the self-adjoint extension
$\tilde T_{\alpha,\beta}$ is unique.  The spectrum is actually given
by \eqref{eq:Jacobidiffeq}.

 In order to prove that the polynomial eigenfunctions $\{ y_n\}_{n=1}^\infty$ are indeed the $X_1$-Jacobi polynomials it is enough to note that both sequences span the same flag of subspaces
  \[\operatorname{span}\{y_1,\dots,y_n \}=\operatorname{span}\{\hPab_1,\dots,\hPab_n \}=\cE^{a,b}_n,\qquad \forall n\geq1\]
and they are orthogonal with respect to the same weight, so up to a multiplicative factor they must coincide.

\subsection*{Converse statement of Theorem \ref{thm:main1}}
By assumption $T$ is a second order differential operator
with a complete set $\{y_n\}_{n=1}^\infty$ of polynomial
eigenfunctions. Without loss of generality we can assume that the
coefficients $p(x)$, $q(x)$ and $r(x)$ are rational functions (see
Proposition 3.1 in \cite{GKMpart1}).

If a constant $y_0$ is a formal eigenfunction of $T$, then Bochner's
theorem implies that $\{y_n\}_{n=0}^\infty$ is one of the classical
orthogonal polynomial systems.  Hence, the sequence starting with
$y_1$ is either not dense (contrary to the assumptions) or constitutes
a classical 1-OPS.

Let us therefore assume that $T$ has polynomial eigenfunctions for all
degrees $n\geq1$ but not for $n=0$. The converse statement of Theorem
\ref{thm:part1main} asserts that $T$ must be of the form
\eqref{eq:TX1def} with \eqref{eq:Tpqrdef}. Up to an affine
transformation of $x$, $p(x)$ assumes one of the following five
canonical forms:
\begin{subequations}
\begin{eqnarray}
\textrm{i)}\qquad p(x)&=& 1-x^2,\\
\textrm{ii)}\qquad p(x)&=& 1+x^2,\\
\textrm{iii)}\qquad p(x)&=& x^2,\\
\textrm{iv)}\qquad p(x)&=& x,\\
\textrm{v)}\qquad p(x)&=& 1,
\end{eqnarray}
\end{subequations}
Writing each of the above 5 cases in self-adjoint form, we
  obtain the following expressions for the weight factor determined by \eqref{eq:Wintform}
\begin{subequations}
\begin{eqnarray}
\textrm{i)}\qquad W(x)&=& \frac{(x-1)^{-a+ab} (x+1)^{a+ab}}{(x-b)^2}\label{eq:W1}\\
\textrm{ii)}\qquad W(x)&=& \frac{{\rm e}^{2a \arctan x} (1+x^2)^{ ab}}{(x-b)^2}\\
\textrm{iii)}\qquad W(x)&=& \frac{x^{2 a b}}{(x-b)^2}\\
\label{eq:W4}
\textrm{iv)}\qquad W(x)&=& \frac{{\rm e}^{ax}x^{ab}}{(x-b)^2}\\
\textrm{v)}\qquad W(x)&=& \frac{{\rm e}^{2 a x}}{(x-b)^2}
\end{eqnarray}
\end{subequations}
Note that the interval cannot include $x=b$ since all eigenpolynomials must be square-integrable. We can then use Green's identity \eqref{eq:Green} where $f$ and $g$ are any linear combination of the polynomial eigenfuncions of $T$. Theorem \ref{thm:part1main} states that the eigenpolynomials span $\cE^{a,b}$, and since
  \[ (x-b)^2 \cP \subset \cE^{a,b}\] then Green's identity \eqref{eq:Green}  holds, in
  particular, for
  \[ f(x) = (x-b)^2 f_1(x),\quad g(x) = (x-b)^2 g_1(x),\]
  where $f_1, g_1$ are arbitrary polynomials. We observe that
  \begin{align*}
     f'(x) g(x) - f(x) g'(x)
    &= (x-b)^4 (f_1'(x) g_1(x) - f_1(x)
    g_1'(x))\\
    &= (x-b)^4 h(x),
  \end{align*}
  where
  \[ h(x) = f_1'(x) g_1(x) - f_1(x) g_1'(x),\]
  is an arbitrary polynomial. Since the left-hand side of \eqref{eq:Green} vanishes by assumption, the
  expression
  \begin{equation}
     \Big[p(x) W(x) (x-b)^4 h(x)\Big]^{x_2}_{x_1} =0
  \end{equation}
  must vanish for all polynomials $h(x)$, which implies that
  \begin{equation}
    \label{eq:pWzeros}
     (pW)(x_1)  = (pW)(x_2)  = 0,
  \end{equation}
where the above evaluations have to be understood in the limit sense if one or both
  of the endpoints $x_1,x_2$ are infinite.

It is clear that condition \eqref{eq:pWzeros} excludes cases (ii)
  and (v) for all possible choices of $x_1, x_2$. Case (iii) is also excluded by the requirement that all eigenpolynomials be square-integrable relative to $W dx$ on $[x_1,x_2]$.

Case (i) leads naturally to the $X_1$-Jacobi SLP. Equation \eqref{eq:pWzeros}
  implies
  \begin{equation}\label{eq:abconds}
    x_1=-1, \quad x_2 = 1,\qquad
    ab\pm a > -1.
  \end{equation}
Setting
  \[ \alpha = ab-a,\quad \beta = ab+a,\] we obtain \eqref{eq:abdef} and the conditions on $\alpha,\beta$
  given at the beginning of Section \ref{sec:X1}. In particular, equation \eqref{eq:abconds} implies  \eqref{alfabetacond1} while \eqref{alfabetacond2} has to be imposed to ensure that $b$ lies outside the interval $[-1,1]$. With these restrictions, the weight \eqref{eq:W1} specializes to the $X_1$-Jacobi weight $\hWab$ shown in \eqref{eq:Wabdef}. Theorem \ref{thm:part1main} implies that the eigenpolynomials of the
 SLP are the $X_1$-Jacobi polynomials.

Similarly, case (iv) corresponds to the $X_1$-Laguerre SLP .  By rescaling $x$ we can assume that $a=-1$ without
  loss of generality. The condition \eqref{eq:pWzeros} implies then
  \begin{gather}
    x_1 = 0,\; x_2 = +\infty,\quad b<1.
  \end{gather}
  However, for $b$ to lie outside $[x_1,x_2]$, we must impose
  $b<0$.  Setting
  \[ a=-1,\quad b=-k\]
  we obtain the $X_1$-Laguerre weight \eqref{eq:Wkdef} by specializing
  the weight shown in \eqref{eq:W4}.  The same argument as above shows
  that the given SLP has to be the $X_1$-Laguerre SLP.
  \qed

%
%

\section{Properties of $X_1$-Jacobi polynomials}\label{sec:propJ}

\subsection{Factorization and Rodrigues formula}
Define the following lowering and raising operators:
\begin{eqnarray}\label{eq:JacA}
  A_{\alpha,\beta}(y) &=&
  \frac{(x-c)}{(x-b)}\,\big(y'+ay\big)-a y,\\
  &=& \frac{(x-c)^2}{x-b}\, \frac{d}{dx} \left(\frac{y}{x-c}\right),\\
   \label{eq:JacB}
  B_{\alpha,\beta}(y) &=&
  (x^2-1)\left(\frac{x-b}{x-c}\right)(y' + ay) -a(x^2-2bx+1)y\\
   \label{eq:JacB1}
  &=& -\big((x-c)\hWab\big)^{-1}
  \frac{d}{dx}\left(\frac{(x-c)^2}{x-b}\hat W_{\alpha+1,\beta+1}\, y\right)
\end{eqnarray}
where $a,b,c$ are related to $\alpha,\beta$ by
\eqref{eq:abcfromalphabeta}, and where the weight $\hWab$
is defined in \eqref{eq:Wabdef}.  Using the above operators we can
factorize $T_{\alpha,\beta}$ in two different ways:
\begin{subequations}\label{eq:Tii}
  \begin{eqnarray}
    T_{\alpha,\beta}&=& B_{\alpha,\beta}\,A_{\alpha,\beta}\\
    &=&  A_{\alpha-1,\beta-1}\,B_{\alpha-1,\beta-1} - \alpha-\beta
    \label{eq:Tii2}.
  \end{eqnarray}
\end{subequations}

\noindent
Another consequence of \eqref{eq:JacA} \eqref{eq:JacB} is the
following adjoint-type relation
\begin{equation}
  (A_{\alpha,\beta}\,f,g)_{\alpha+1,\beta+1}   (f,B_{\alpha,\beta}\,g)_{\alpha,\beta}.
\end{equation}relative to the the inner products
defined in \eqref{eq:Wabdef}\eqref{eq:jacobiproduct}.

By virtue of  the intertwining relations
\eqref{eq:Tii}, the above raising operator can be applied iteratively
to construct the $X_1$-Jacobi polynomials. The difference with
respect to the classical raising operators is that on each iteration a
parameter needs to be shifted by an additive constant. More
specifically, the following relations hold:
\begin{eqnarray}
\label{eq:Aabaction}
&& A_{\alpha,\beta}\hPab_{n}\frac{1}{2}(n+\alpha+\beta)\,\hP{\alpha+1,\beta+1}_{n-1}\\
\label{eq:Babaction}
&& B_{\alpha,\beta}\hP{\alpha+1,\beta+1}_n= 2n\,\hPab_{n+1}
\end{eqnarray}
where $\hPab_{n}$ is the $n$-th $X_1$-Jacobi polynomial.

For fixed $\alpha,\beta$, set
\begin{eqnarray}
   b_j &=& b+j/a,\\
   \hat W_j &=& \hat W_{\alpha+j,\beta+j},\\
   \tilde{B}_j(y)&=&\frac{d}{dx}\left[
  \frac{(x-b_j)^2}{(x-b_{j-1})(x-b_{j+1})}\, y\right].
\end{eqnarray}
Iterating \eqref{eq:Babaction} and using the identity
\[
\hP{\alpha+j,\beta+j}_1=-\frac{1}{2}(x-b_{j+1}),\] we obtain the
following Rodrigues-type formula for the $X_1$-Jacobi polynomials:
\begin{equation}
 (-2)^n (n-1)! \hP{\alpha,\beta}_n=\frac{(\tilde{B}_1 \cdots
   \tilde{B}_{n-1})
    \big[(x-b_{n})^2 \hat W_{n-1}\big]}{(x-b_1) \hat W_0}
\end{equation}


\subsection{Norms}

The square of the norm of the $X_1$-Jacobi polynomials is given by
\begin{equation}
  \label{eq:JacL2X1}
  \small
  \int_{-1}^1 \frac{(1-x)^\alpha (1+x)^\beta}{(x-b)^2} \left(
    \hPab_n\right)^2 \, dx = \frac{(\alpha+n) (\beta+n)}{4
    (\alpha+n-1)(\beta+n-1)}\, C_{n-1},
\end{equation}
where
\begin{equation}
  \label{eq:Cndef}
  C_n
  = \frac{2^{\alpha+\beta+1}}{\alpha+\beta+2n+1}\,
  \frac{\Gamma(\alpha+n+1) \Gamma(\beta+n+1)}{\Gamma(n+1)
    \Gamma(\alpha+\beta+n+1)}.
\end{equation}
The above should be contrasted with the norm formula for the classical
Jacobi polynomials, namely:
\begin{equation}
  \label{eq:JacL2ord}
  \int_{-1}^1 (1-x)^\alpha (1+x)^\beta \left(
    P^{(\alpha,\beta)}_n\right)^2 \, dx =  C_n.
\end{equation}
\subsection{Relation to classical polynomials}
The $X_1$-Jacobi polynomials $\hPab_n$ are related to their classical
counterparts $P^{(\alpha,\beta)}_n$ by the following 3-term linear combination:
\begin{equation}
  \label{eq:JacX1ord0}
  \hPab_n = -\frac{1}{2}(x-b)P^{(\alpha,\beta)}_{n-1}+
  \frac{b P^{(\alpha,\beta)}_{n-1} -
    P^{(\alpha,\beta)}_{n-2}}{(\alpha+\beta+2n-2)},
\end{equation}
where $b$ is given by \eqref{eq:abcfromalphabeta}.  Using the 3-term
recurrence relation for the classical Jacobi polynomials,
relation \eqref{eq:JacX1ord0} can be rewritten as
\begin{equation}
  \label{eq:JacX1ord}
  \hPab_n =  -f_n \, P^{(\alpha,\beta)}_n  +2\,b\,g_n\,
  P^{(\alpha,\beta)}_{n-1} - h_n\, P^{(\alpha,\beta)}_{n-2},
\end{equation}
where
\begin{eqnarray}
  f_n &=& \frac{n\,
    (\alpha+\beta+n)}{(\alpha+\beta+2n-1)(\alpha+\beta+2n)},\\
  g_n &=&
  \frac{(\alpha+n)(\beta+n)}{(\alpha+\beta+2n-2)(\alpha+\beta+2n)},\\
    h_n &=&
    \frac{(\alpha+n)(\beta+n)}{(\alpha+\beta+2n-2)(\alpha+\beta+2n-1)},
\end{eqnarray}
and where $b$ is given by \eqref{eq:abcfromalphabeta}.
Relation \eqref{eq:JacX1ord} can be established by means of
\eqref{eq:Aabaction} and by using the series definition of the
classical Jacobi polynomials.  The details are left to the reader.
Using \eqref{eq:JacL2X1} \eqref{eq:JacL2ord} and the orthogonality
properties of $\hPab_n, P^{(\alpha,\beta)}_n$, relation \eqref{eq:JacX1ord} can
be inverted to obtain the following identity:
\begin{equation}
  \label{eq:JacX1ordreverse}
  -\frac{1}{4}(x-b)^2\, P^{(\alpha,\beta)}_n = f_{n+1} \hPab_{n+2} -2\,b\,
 \hat{g}_n
  \hPab_{n+1} + \hat{h}_n \hPab_n,
\end{equation}
where
\begin{eqnarray}
  \hat{g}_n &=&   \frac{(n+\alpha)(n+\beta)}{(\alpha+\beta +
    2n)(\alpha+\beta+2n+2)},\\
  \hat{h}_n &=&   \frac{(n-1+\alpha)(n-1+\beta)}{(\alpha+\beta +
    2n)(\alpha+\beta+2n+1)}.
\end{eqnarray}

\subsection{Recursion formula}
Using \eqref{eq:JacX1ord} and the classical 3-term recurrence identity
we obtain the following expression for the classical Jacobi polynomial
in terms of  its $X_1$ counterparts:
\begin{eqnarray}
  \frac{1}{4}(b^2-1)P^{(\alpha,\beta)}_n &=& (\alpha+n)(\beta+n)\left(-f_{n+1}
  \hPa_{n+2}+\frac{x}{2}\,\hPa_{n+1}\right)\\ \nonumber
  & -&2(a^2-1)b\,\hat{g}_n\, \hPab_{n+1}\\ \nonumber
  &-& (\alpha+n+1)(\beta+n+1)\,\hat{h}_n\, \hPa_n,
\end{eqnarray}
where $a,b$ are given by \eqref{eq:abcfromalphabeta}.
Combining the above identity with \eqref{eq:JacX1ordreverse} yields
the following 3-term recurrence for the $X_1$-Jacobi polynomials:
\begin{eqnarray}
  f_{n+1}[(b^2-1)-(\alpha+n)(\beta+n)(x-b)^2]\,\hPab_{n+2} &+&\\ \nonumber
  -2b\, \hat{g}_n \big[(b^2-1)+(a^2-1)(x-b)^2\big] \hPab_{n+1}&+&\\ \nonumber
    \frac{1}{2}(\alpha+n)(\beta+n)x(x-b)^2\hPab_{n+1}&+&\\ \nonumber
    \hat{h}_n \big[(b^2-1)-(\alpha+n+1)(\beta+n+1)(x-b)^2\big]\,\hPab_n
    &=& 0.
\end{eqnarray}



\subsection{First few $X_1$-Jacobi polynomials}
The first few $\hPab_n$ polynomials are:
\begin{subequations}
\begin{eqnarray*}
&&\hPab_1=-\tfrac{1}{2}\,x-\tfrac{2+\alpha +\beta }{2 (\alpha -\beta )},\\
&&\hPab_2=-\tfrac{\alpha +\beta+2 }{4}\, x^2 -\tfrac{  \alpha ^2+\beta ^2 +2 (\alpha+\beta) }{2 (\alpha -\beta )}\,x-\tfrac{\alpha+\beta+2}{4},\\
&&\hPab_3=-\tfrac{(\alpha +\beta+3)(\alpha +\beta+4 )}{16} \,x^3  -\tfrac{ (3+\alpha +\beta ) \left(6 \alpha +3 \alpha ^2+6 \beta -2 \alpha  \beta +3 \beta ^2\right)}{16 (\alpha -\beta )}\,x^2\\
&&\qquad\qquad-\tfrac{\left(9 \alpha +3 \alpha ^2+9 \beta +2 \alpha
\beta +3 \beta ^2\right)}{16}\, xs -\tfrac{-6 \alpha +\alpha
^2+\alpha ^3-6 \beta -6 \alpha  \beta -\alpha ^2 \beta +\beta
^2-\alpha  \beta ^2+\beta ^3}{16 (\alpha -\beta )},
\end{eqnarray*}
\end{subequations}
\subsection{Zeroes of $X_1$-Jacobi polynomials}

Many properties of the zeros of the $X_1$-Jacobi polynomials follow from the fact that they are  eigenfunctions of a Sturm-Liouville problem.
However, we choose to give a direct proof below independent of Sturm-Liouville theory.

\begin{prop}\label{prop:x1jroots}
Assume without loss of generality that $a<0$, then the $n$-th Jacobi polynomial $\hPab_n(x)$ has one zero in $(-\infty,b)$ and $n-1$ zeroes in $(-1,1)$.
\end{prop}

Before proving Proposition \ref{prop:x1jroots}, let us state the following two lemmas:

\begin{lem}\label{lem1}
  Let $P \in\cE^{a,b}_n$ be a polynomial with $n$ real roots.
  If $a<0$ and $P(b)\neq 0$, at least one of these roots lies in $(-\infty,b)$.
\end{lem}
\begin{proof}
  From \eqref{eq:Eabdef} it follows that
  \[ a = -P'(b)/P(b),\]  hence $P(b)$ and
  $P'(b)$ have the same sign. By Sturm's root counting theorem, It is clear that a root of $P$ has to lie in $(-\infty,b)$ otherwise $P$ cannot have $n$ real roots.
\end{proof}

\begin{lem}\label{lem2}
 $\hPab_n(b)\neq 0$.
\end{lem}
\begin{proof}
 First note that the $\hPab_n(x)$ are defined recursively by \eqref{eq:Babaction} and \eqref{eq:JacB1}. Using \eqref{eq:Wabdef} in \eqref{eq:JacB1} it is clear that \eqref{eq:JacB1} has the form
 \[ B_{\alpha,\beta} y=(x-b)^2 f(x) \frac{d}{dx}\left(\frac{g(x)}{(x-b)} y\right), \]
 where $f(b)\neq0$ and $g(b)\neq0$. Since $\hPab_1(b)\neq0$, it follows by induction that $\hPab_n(b)\neq0$ for all $n>1$.

\end{proof}

\begin{proof}[Proof of Proposition \ref{prop:x1jroots}]
  From the two previous lemmas
   it follows that $\hPab_{n}(x)$
  has at most $n-1$ zeroes in $(b,\infty)$, and in particular at most
  $n-1$ zeroes in  $(-1,1)$.  Suppose that $\hPab_{n}(x)$ has
  $\xi_1,\ldots, \xi_k$, $1\leq k\leq n-2$  zeroes in $(-1,1)$, and let
    \[ Q_1(x) := (x-\xi_1)\cdots (x-\xi_k).\]
    If $\hPab_{n}(x)$ has no zeros in $(-1,1)$ then take $Q_1(x)=1$.
  By Lemma \ref{lem1}, the polynomial $Q_1\notin\cE^{a,b}(x)$ but we can always choose $\xi$ so that
  \[ Q(x) := (x-\xi) Q_1(x)\in \cE^{a,b}_{n-1}.\]
 This is clear because imposing \eqref{eq:Eabdef} on the above expression leads to
  \[  (b-\xi) (Q_1'(b) + a Q_1(b)) + Q_1(b) = 0,\]
  which can be solved for $\xi$.  Again, Lemma \ref{lem1} implies that $\xi\notin
  (-1,1)$, and therefore the function $Q(x)\hPab_n(x)$ does not change sign for
  $x\in[-1,1]$. Hence
  \[ \left( \hPab_n, Q\right)_{\alpha,\beta} \neq0.\]
but this is impossible since $\hPab_n$ is orthogonal
  to $\cE^{a,b}_{n-1}$.  We conclude then that $\hPab_n$ has exactly $n-1$ roots in
  $(-1,1)$. The remaining root has to be real and Lemma \ref{lem1} implies that it lies in $(-\infty,b)$.
\end{proof}

\section{Properties of $X_1$-Laguerre  polynomials}\label{sec:propL}

\subsection{Factorization and Rodrigues formula}
Define the following lowering and raising operators:
\begin{eqnarray}
  \label{eq:LagA}
  A_k(y) &=&  -\frac{(x+k+1)}{(x+k)}\,\big(y'-y\big)- y,\\
  &=& \frac{(x+k+1)^2}{x+k}\frac{d}{dx} \left[ \frac{y}{x+k+1}\right]\\
   \label{eq:LagB}
  B_k(y) &=&
  x\,\frac{(x+k)}{(x+k+1)}\,\left(y' -y\right) +k y\\
  \label{eq:LagB1}
  &=& \big( (x+k+1)\hat W_k \big)^{-1} \frac{d}{dx}\left[\frac{(x+k+1)^2}{x+k}
  \hat W_{k+1}\, y\right],
\end{eqnarray}
where the weight $\hat W_k$ is defined in \eqref{eq:Wkdef}.
Note that we can factorize the second-order operator in
\eqref{eq:Tkdef} in two different ways:
\begin{subequations}\label{eq:Ti}
\begin{eqnarray}\label{eq:Ti1}
T_k&=& B_k\,A_k\\
&=& A_{k-1} B_{k-1} -1, \label{eq:Ti2}
\end{eqnarray}
\end{subequations}
and observe the
following relations relative to the the inner products
defined in \eqref{eq:Wkdef}\eqref{eq:Laguerreproduct}:
\begin{equation}
  \label{eq:LagX1adjoint}
  (A_k\,f,g)_{k+1} = (f,B_k\,g)_k.
\end{equation}
By virtue of the intertwining relations
\eqref{eq:Ti}, the raising operator $B_k$ can be applied iteratively
to construct the $X_1$-Laguerre polynomials. The difference with
respect to the classical raising operators is that on each iteration a
parameter needs to be shifted by an additive constant. More
specifically, the following relations hold:
\begin{eqnarray}
\label{eq:Akaction}
&& A_k\hL{k}_{n}= \hL{k+1}_{n-1}\\
\label{eq:Bkaction}
&& B_k\hL{k+1}_n= n\,\hL{k}_{n+1}
\end{eqnarray}
where $\hL{k}_{n}$ is the $n$-th $X_1$-Laguerre polynomial.
Iterating \eqref{eq:Bkaction} we obtain
\begin{eqnarray}
  \label{eq:LagRod1}
  (n-1)!\,\hL{k}_n &=& (B_k\cdots B_{k+n-2})\hL{k+n-1}_1,\qquad n=2,3,\dots
\end{eqnarray}
Fix $k$ and set
\begin{eqnarray}
\tilde{B}_j(y)&=&\frac{d}{dx}\left[
  \frac{(x+k+j)^2}{(x+k+j-1)(x+k+j+1)}\,
  y\right].
\end{eqnarray}
Using \eqref{eq:LagB1} and
\[ \hL{k}_1=-(x+k+1),\] we rewrite \eqref{eq:LagRod1} to obtain the
following Rodrigues-type formula for the $X_1$-Laguerre
polynomials:
\begin{equation}
-(n-1)!\,\hL{k}_n=\frac{(\tilde{B}_1 \cdots
\tilde{B}_{n-1}) \big[(x+k+n)^2 \hat W_{k+n-1}\big]}{(x+k+1)\hat W_k}
\end{equation}

\subsection{Norms}

The square of the norm of the $X_1$-Laguerre polynomials is given by
\begin{equation}
  \label{eq:X1laguerreL2}
  \big(\hL{k}_n\big)^2\, dx
  =\frac{(k+n-1)}{(k+n)}K_{n-1}
\end{equation}
The above relation follows by induction from \eqref{eq:Laguerrediffeq}
\eqref{eq:Ti1} \eqref{eq:LagX1adjoint} \eqref{eq:Akaction}.  Contrast
the above to the norm formula for the classical Laguerre polynomials,
namely:
\begin{equation}
  \label{eq:laguerreL2}
  \int^\infty_0 x^k e^{-x} \big(L^{(k)}_n\big)^2\, dx   \frac{\Gamma(n+k+1)}{n!}\equiv K_n
\end{equation}

\subsection{Relation to classical polynomials}
The $X_1$-Laguerre polynomials $\hLk_n$ are related to the classical
Laguerre polynomials $L^{(k)}_n$ by the following simple relation:
\begin{equation}
  \label{eq:LagX1ord0}
    \hLk_n = -(x+k+1) L^{(k)}_{n-1}+L^{(k)}_{n-2}.
\end{equation}
Using the 3-term recurrence relation for the classical $L^{(k)}_n$,
\begin{equation}
    \label{eq:LagRec}
    n L^{(k)}_n +(x-2n-k+1)L^{(k)}_{n-1}
    +(n+k-1)L^{(k)}_{n-2}=0,
\end{equation}
relation \eqref{eq:LagX1ord0} may be rewritten as
\begin{equation}
  \label{eq:LagX1ord}
  \hLk_n = n L^{(k)}_n-2(n+k)L^{(k)}_{n-1} + (n+k)L^{(k)}_{n-2}.
\end{equation}
The above identity follows by induction from \eqref{eq:Akaction} and
from the following properties of the classical Laguerre polynomials:
\begin{eqnarray}
  &&L^{(k)}_n = L^{(k+1)}_n - L^{(k+1)}_{n-1},\\
  && \frac{dL^{(k)}_n}{dx} = -L^{(k+1)}_{n-1}.
\end{eqnarray}

Using \eqref{eq:X1laguerreL2} \eqref{eq:laguerreL2} and the
orthogonality properties of $\hLk_n$ and $L^{(k)}_n$ we can invert
\eqref{eq:LagX1ord} to obtain the following identity:
\begin{equation}
  \label{eq:LagX1ordreverse}
  (x+k)^2 L^{(k)}_n = (n+1) \hLk_{n+2}-2(n+k)\hLk_{n+1}+(n+k-1)\hLk_n.
\end{equation}
\subsection{Recursion formula}
Identities \eqref{eq:LagRec}  and  \eqref{eq:LagX1ord}
imply the following:
\begin{eqnarray}
  \label{eq:LagX1ordreverse1}
  (n+1) (n+k)\, \hLk_{n+2}&+&\\\nonumber
  +(n+k)(x-2n-k-1)\, \hLk_{n+1} &+&\\\nonumber
  +(n+k-1)(n+k+1) \,  \hLk_{n} &=& k L^{(k)}_n.
\end{eqnarray}
Combining this with \eqref{eq:LagX1ordreverse} yields the following
3-term recurrence for the $X_1$-Laguerre polynomials:
\begin{eqnarray}
  \label{eq:LagX1Rec}
  (n+1) [(x+k)^2(n+k)-k]\, \hLk_{n+2}&+&\\ \nonumber
  +(n+k)[(x+k)^2(z-2n-k-1)+2k]\, \hLk_{n+1} &+&\\ \nonumber
  +(n+k-1)[(x+k)^2(n+k+1)-k] \,  \hLk_{n} &=& 0.
\end{eqnarray}
\subsection{Zeroes of $X_1$-Laguerre polynomials}

\begin{prop}\label{prop:x1lroots}
The $n$-th Laguerre polynomial $\hLk_n(x)$ has one zero in $(-\infty,-k)$ and $n-1$ zeroes in $[0,\infty)$.
\end{prop}
\begin{proof}
The proof follows the same arguments as Proposition \ref{prop:x1lroots}.
\end{proof}

\subsection{First few $X_1$-Laguerre polynomials}
The first few $\hLk_n$ polynomials are:
\begin{subequations}
\begin{eqnarray}
\hLk_1&=&-x-(1+k),\\
\hLk_2&=&x^2-k(k+2),\\
\small \hLk_3&=&-\tfrac{1}{2}x^3+\tfrac{k+3}{2} x^2
+\tfrac{k(k+3)}{2} x -\tfrac{k}{2} \left(3  +4 k +k ^2\right).
\end{eqnarray}
\end{subequations}

\vskip0.5cm
\paragraph{\textbf{Acknowledgments}}
\thanks{ We are grateful to Jorge Arves\'u, Mourad Ismail, Francisco
  Marcell\'an and Andr\'e Ronveaux for their helpful comments. A
  special note of thanks goes to Norrie Everitt for his suggestions
  and remarks regarding operator domains and the limit point/circle
  analysis, and to Lance Littlejohn for comments regarding classical
  polynomials with negative integer parameters.
  The research of DGU is supported in
  part by the Ram\'on y Cajal program of the Spanish ministry of
  Science and Technology and by the DGI under grants MTM2006-00478 and
  MTM2006-14603. The research of NK is supported in part by NSERC
  grant RGPIN 105490-2004. The research of RM is supported in part by
  NSERC grant RGPIN-228057-2004. }


\begin{thebibliography}{99}


\bibitem{Aczel} J. Aczel, Eine Bemerkung \"uber die Charakterisierung der klassichen orthogonale Polynome, \emph{Acta Math. Acad.Sci. Hungar} \textbf{4} (1953), 315-321.


\bibitem{amr02} M. Alfaro, M. \'{A}lvarez de Morales, M. L. Rezola,
  Orthogonality of the Jacobi polynomials with negative integer
  parameters, \textit{ Journal of Computational and Applied
    Mathematics}, \textbf{145} (2002) 379--386.

\bibitem{askey} R.A. Askey and J.A. Wilson, Some basic hypergeometric orthogonal polynomials that generalize Jacobi polynomials, \emph{Memoirs American Mathematical Society} No. \textbf{319} (1985).


\bibitem{atkinson} F.V. Atkinson and W.N. Everitt, Orthogonal polynomials which satisfy second order differential
 equations. E. B. Christoffel (Aachen/Monschau, 1979), pp. 173--181, Birkhäuser, Basel-Boston, Mass.,  1981.


%
\bibitem{Bo} S. Bochner, \"Uber Strum-Liouvillsche Polynomsysteme, \emph{Math. Z.} \textbf{29} (1929), 730-736.


\bibitem{elw04} W. N.  Everitt, L. L.  Littlejohn, R.  Wellman, The
  Sobolev orthogonality and spectral analysis of the Laguerre
  polynomials ${L\sp {-k}\sb n}$ for positive integers $k$, \textit{J.
    Comput. Appl. Math.}  \textbf{171} (2004), 199--234.


\bibitem{EverittJL} W. N. Everitt, Note on the X1-Jacobi orthogonal
  polynomials, \texttt{arXiV CA 0812.0728} and Note on the
  X1-Laguerre orthogonal polynomials, \texttt{arXiV CA 0812.3559}

\bibitem{EKLW} W. N. Everitt, K. H. Kwon, L. L. Littlejohn and R. Wellman, Orthogonal polynomial solutions of
linear ordinary differential equations, \emph{J. Comp. Appl. Math} \textbf{133} (2001), 85–109.


\bibitem{Feldmann} J. Feldmann, On a characterization of classical orthogonal polynomials, \emph{Acta. Sc. Math.}
\textbf{17} (1956), 129–133.

%
%

\bibitem{GKMpart1}
D.~G{\'o}mez-Ullate, N.~Kamran, and R.~Milson,
\newblock An extension of Bochner's problem: exceptional invariant subspaces
\texttt{arXiV math-ph 0805.3376}


\bibitem{grunbaum} A. Gr\"unbaum and L. Haine,  The $q$-version of a theorem of Bochner, \emph{J. Comput. Appl. Math.} \textbf{68} (1996), 103--114.

\bibitem{He} E. Heine, \emph{Theorie der Kugelfunctionen und der verwandten Functionen}, Berlin, 1878.


\bibitem{hendrikssen} E. Hendriksen and H. van Rossum,  Semiclassical orthogonal polynomials, in
 ``Orthogonal polynomials and applications'' (Bar-le-Duc, 1984),
 354--361, \emph{Lecture Notes in Math.}, \textbf{1171}, Springer, Berlin,  1985.

\bibitem{ismail} M.E.H. Ismail , A generalization of a theorem of Bochner, \emph{J. Comp. Appl. Math.}
 \textbf{159} (2003), 319--324.

\bibitem{IsvAs} M.E.H. Ismail and W. van Assche, \emph{Classical and quantum orthogonal polynomials in one variable}, Encyclopedia in Mathematics, Cambridge University Press, Cambridge, 2005.

\bibitem{Krall} H. L. Krall, On orthogonal polynomials satisfying a certain fourth order differential equation, The
Pennsylvania State College Studies, No 6, 1940.


\bibitem{KL97}  K. H. Kwon and L. L. Littlejohn, Classification of classical orthogonal polynomials, \emph{J. Korean Math. Soc.} \textbf{34} (1997), 973--1008.

\bibitem{Lesky} P. Lesky, Die Charakterisierung der klassischen
  orthogonalen Polynome durch SturmLiouvillesche Differentialgleichungen,\emph{ Arch. Rat. Mech. Anal.} \textbf{10}
  (1962), 341--352.

\bibitem{Mikolas}  M. Mikol\'as, Common characterization of the Jacobi, Laguerre and Hermite-like
polynomials(in Hungarian), \emph{Mate. Lapok} \textbf{7} (1956), 238--248.

\bibitem{Q} C. Quesne, Exceptional orthogonal polynomials, exactly solvable potentials and
supersymmetry, \emph{J. Phys. A: Math. Theor.} \textbf{41} No. 39,
392001.

\bibitem{routh} E.J. Routh, On some properties of certain solutions of a
  differential equation of the second order, \textit{Proc. London Math. Soc.},
  \textbf{16} (1885), 245-261.

\bibitem{RS}  M.~Reed and B.~Simon, {\it Methods of Modern Mathematical Physics. II.
Fourier Analysis, Self-Adjointness}, Academic Press, New York-London, 1975.

\bibitem{Ron87} A. Ronveaux, Sur l'\'equation diff\'erentielle du second ordre satisfaite par une
 classe de polyn\^omes orthogonaux semi-classiques,  \emph{C. R. Acad. Sci. Paris S\'er. I Math.}  \textbf{305}  (1987),  no. 5, 163--166.

\bibitem{Ron79} A. Ronveaux, Polyn\^omes orthogonaux dont les polyn\^omes d\'eriv\'es
 sont quasi orthogonaux,  \emph{C. R. Acad. Sci. Paris S\'er. A-B}  \textbf{289}  (1979),  no. 7, A433--A436.

\bibitem{RM89} A. Ronveaux and F.  Marcell\'an,  Differential equation for classical-type orthogonal polynomials,
\emph{ Canad. Math. Bull.}  \textbf{32}  (1989),  no. 4, 404--411.

\bibitem{St} T.J. Stieltjes, Sur certains polyn\^omes qui
v\'erifient une \'equation diff\'erentielle du second ordre et sur la th\'eorie des
fonctions de Lam\'e, \emph{Acta Mathematica} \textbf{6} (1885), 321-326.

\bibitem{Sz} G. Szeg\"o, \textit{Orthogonal polynomials}, Colloquium
Publications \textbf{23}, American Mathematical Society, Providence, 1939.

\bibitem{U} V. B. Uvarov, The connection between systems of polynomials that
are orthogonal with respect to different distribution functions,
\emph {USSR Computat. Math. and Math. Phys.} \textbf{9} (1969),
25--36.

\end{thebibliography}
\end{document}